\documentclass[aps,preprint]{revtex4}
\usepackage{graphicx}
\begin{document}
\title{\Large \bf SUSY in the Spacetime of Higher-Dimensional Rotating Black Holes}
\author{\large Haji Ahmedov and Alikram N. Aliev}
\address{Feza G\"ursey Institute, P. K. 6  \c Cengelk\" oy, 34684 Istanbul, Turkey}
\date{\today}

\begin{abstract}
General higher-dimensional rotating black hole spacetimes of any  dimensions admit the Killing and Killing-Yano tensors, which generate the hidden symmetries just as in four-dimensional Kerr spacetime. We study these properties of the black holes using the formalism  of supersymmetric mechanics of  pseudo-classical spinning point particles. We present two nontrivial supercharges, corresponding to the  Killing-Yano and conformal Killing-Yano tensors of the second rank. We demonstrate that  an unusual extended Poisson-Dirac algebra of these supercharges  results in two independent Killing tensors in spacetime dimensions $ D\geq 6 $, giving explicit examples for the Myers-Perry black holes in  $ D = 6 $ dimensions.
\end{abstract}


\maketitle

\section{Introduction}

Black holes are supposed to be one of the most enigmatic objects in nature and it is remarkable that  general relativity provides an exact mathematical description  of these objects. As is known, the exact solution of the Einstein  field equations discovered  by R. Kerr in 1963  describes  a family of rotating black holes \cite{kerr}. The Kerr metric  is stationary and axisymmetric, that implies  its invariance under two continuous spacetime symmetries: time-translational and rotational symmetries defined by two commuting  Killing vector fields. Another important symmetry property of the Kerr metric was revealed when exploring  its geodesics. In 1968, Carter \cite{carter1} showed that the Hamilton-Jacobi equation for geodesics of the Kerr metric admits a complete separation of variables. The reason for this was the existence of an extra constant of motion not related to the global isometries of the spacetime. He was also able to show the separability of variables in the Klein-Gordon equation for charged particles by constructing explicitly the set of four, mutually commuting, differential operators  \cite{carter2}. In 1970, Walker and Penrose  gave \cite{wp} an elegant mathematical  interpretation of these results, pointing out that the Kerr metric admits hidden symmetries  generated by a second rank Killing tensor. Namely, the Killing tensor plays a crucial role in the complete integrability of the geodesic and  scalar field equations.

The existence of the Killing tensor has also motivated  the study of electromagnetic and gravitational perturbations of the Kerr spacetime. In 1972, Teukolsky \cite{teuk}  showed that separation of variables occurs in equations  for electromagnetic and gravitational perturbations  and presented a master equation governing scalar, electromagnetic and gravitational perturbations of the Kerr spacetime. A further important step  in the problem of separability in this background  was made by achieving separation of variables in the Dirac equation \cite{chandra1, guven}. In order to explain this result, Carter and Mclenaghan \cite{cartermc}  managed to construct a new linear differential operator, commuting with the Dirac operator. For this purpose, the authors used the fact that the Kerr metric, in addition to the Killing tensor, also admits a second rank antisymmetric, the so-called Killing-Yano  tensor \cite{penrose}. In other words, it is the Killing-Yano  tensor that
provides the separability of variables in the Dirac equation
and its very existence is the physical reason for many remarkable properties of the Kerr metric. These  properties turned out to be  so impressive that Chandrasekhar called them  ``miraculous" properties \cite{chandra2}. The miraculous properties  played a profound  role in astrophysical implications of black holes, facilitating  analytical studies of  various classical and quantum processes  around them.

A systematic exploration of  hidden symmetries generated by the Killing-Yano tensor  was undertaken in a remarkable paper  by Gibbons {\it et al} \cite{gibbons}.  In the strive to answer to mutually correlated  questions of what is the classical analogue of the Carter and Mclenaghan  result and  what is the relation between the Killing-Yano tensor and  the ``fermionic constituent" of point particle dynamics, the  authors  explored the worldline supersymmetric  mechanics of pseudo-classical spinning point particles in curved backgrounds \cite{brink1, ber, brink2, holten}. They found that the existence of the Killing-Yano tensor in the Kerr metric corresponds to the appearance of a {\it new supersymmetry} in   the theory of  spinning point particles in this background.

In recent developments, the efforts in the study of the hidden symmetries of black holes have been focused on higher dimensions. It turned that the hidden symmetries  of the Kerr metric survive in higher dimensions as  well. In 2007, Frolov and  Kubiz\v{n}\`{a}k demonstrated that the higher-dimensional  Myers-Perry metric \cite{mp}, which is a generalization of the Kerr metric to all spacetime dimensions, admits the Killing and Killing-Yano tensors \cite{fk1}. The authors \cite{fk2} have also extended this result to the case of higher-dimensional Kerr-NUT-AdS spacetimes  discovered by  Chen,  L\"u  and  Pope \cite{clp}, which has  subsequently  been the subject for many other studies (see a review paper \cite{fk3}). The existence of the Killing-Yano tensor in the higher-dimensional black hole spacetimes gave a new impetus to the study of the separability of variables for the Dirac equation in these backgrounds \cite{oota, wu1, wu2}. The hidden symmetries of  rotating charged black holes in higher dimensions have been studied in \cite{dkl, ad, chow, kenta}.

In this paper, we explore the hidden symmetries of the general higher-dimensional black hole spacetimes from the point of view of worldline supersymmetric mechanics of pseudo-classical spinning point particles. Following  the work of \cite{gibbons}, we show that the  hidden symmetries of the black hole spacetimes enhance  {\it generic} worldline supersymmetry for  the spinning particles in these spacetimes. We begin with a brief review of  the formalism  of  spinning point particles in curved backgrounds. Next, we consider  two {\it nongeneric}  supercharges, corresponding  to the Killing-Yano and  conformal (closed) Killing-Yano tensors of the second rank and underlying  the extension of the usual worldline supersymmetry.
We show that these supercharges and the standard supercharge of supersymmetric mechanics of the spinning particles are mutually commuting in the sense of Poisson-Dirac brackets. However, the Poisson-Dirac bracket of each of the nongeneric  supercharges with itself  does not  close on the Hamiltonian, forming an unusual algebra. We demonstrate that  this gives rise to two independent  Killing tensors in spacetime dimensions $ D\geq 6 $.

\section{The formalism of Spinning particles}

This formalism is based on the use of anticommuting Grassmann variables to introduce spin degrees of freedom in relativistic mechanics  of point particles \cite{brink1, ber, brink2, holten}. In this sense, it is a pseudo-classical description of the relativistic Dirac particles.  The history of a point particle in ordinary   relativistic mechanics is described by its position vector $ x ^{\mu}( \tau) $ (a Grassmann-even variable) in a spacetime, where $ \tau $ is a proper time parameter along the particle worldline. The extension of the configuration space of this particle by adding a Grassmann-odd  variable $ \psi ^{\mu}(\tau)$, allows one to describe its spin degrees of freedom as well. Remarkably, this also results in the existence of supersymmetry  relating these two variables $ x ^{\mu}( \tau) $ and $ \psi ^{\mu}( \tau) $. Below we  present the action and general relations between the symmetries and constant of motions in  supersymmetric mechanics of the pseudo-classical point particle.

\subsection{The action}

The action for a spinning particle in a $ D $-dimensional curved spacetime with metric $ g_{\mu\nu}(x) $  can be written in the form \cite{brink1}
\begin{equation}
S=\frac{1}{2}\int d\tau \left(e^{-1}\,g_{\mu\nu}\dot{x}^\mu\dot{x}^\nu
    +i\,\psi_a \frac{D\psi^a}{D\tau} +i e^{-1} \chi \psi_a e^{\,a}_\mu\,
    \dot{x}^\mu\right)\,,
    \label{action}
\end{equation}
where $ e(\tau) $ is  an `einbein' field  of the one-dimensional metric of the worldline,  $ e^{\,a}_\mu(x) $ a `vielbein' field  $ e^{\,a}_\mu(x) $  of the spacetime metric and $ \chi $  is its fermionic  counterpart. Thus, we have
\begin{eqnarray}
g_{\mu\nu}& = & \eta_{ab} e^{\,a}_\mu \,e^{\,b}_\nu\,,~~~~~~\psi^a= e^a_ {\,\mu} \,\psi^\mu\,,~~~~~~~~~ \mu\,,\, a= 0,1,...,D\,.
 \label{viels}
\end{eqnarray}
where the indices   $ a $ and  $ b $  are the locally  flat indices and  $ \eta_{ab} $  is a flat Minkowski  metric. Furthermore, the overdot means the usual derivative $ d/d\tau $  and  the  covariant derivative is given by
\begin{equation}
\frac{D\psi^a}{D\tau}= \dot{\psi}^a  -\dot{x}^\mu \,\omega^a_{\,\mu \, b}\,\psi^b\,,
\end{equation}
where $ \omega^a_{\,\mu \, b} $ is the spin connection.  The action (\ref{action}) in its present form  is invariant  with respect to  both a reparametrization  of  $ \tau \rightarrow \tau^\prime (\tau ) $
and local  supersymmetry transformations which are given by
\begin{eqnarray}
 \delta x^\mu &=& -i \epsilon e^\mu_{\,a} \psi^a\,,~~~~\delta e= - 2i \epsilon  \chi\,,~~~~ \delta \chi =  \dot{\epsilon}\,, \nonumber \\[2mm]
\delta \psi^a  &=& \epsilon e^{-1} e^a_{\,\mu} \dot{x}^\mu +\delta
  x^\mu \omega^{\ a}_{\mu\  b}\,\psi^b + i\epsilon e \chi \psi^a\,,
\label{supertrans}
\end{eqnarray}
where $ \epsilon(\tau) $ is an infinitesimal Grassmann-odd parameter.
It is  important to note  the use of these transformations  along with those of the reparametrization invariance  enables one to  make a gauge choice $ e=1 $ and $ \chi=0 $. With this gauge, the action (\ref{action}) reduces to the worldline supersymmetric one of the form \cite{gibbons}
\begin{equation}
S=\int d\tau \left(\frac{1}{2}\,g_{\mu\nu}\dot{x}^\mu\dot{x}^\nu
    +\frac{i}{2}\,\psi_a \frac{D\psi^a}{D\tau} \right)\,.
\label{action1}
\end{equation}
Another important feature of the action (\ref{action}) is that one can consistently derive the desired constraint equations, accompanying  the  usual equations of motion
\begin{eqnarray}
\frac{D^2 x^{\mu}}{D\tau^2}& = &  - R^{\mu}_{\,\,\nu}\, \dot{x}^\nu\,,~~~~~~\frac{D\psi^a}{D\tau} =  0\,,
\label{eqnsmot}
\end{eqnarray}
where
\begin{equation}
R_{\mu\nu}=\frac{i}{2}\,\psi^a \psi^b R_{a b\,\mu\nu}\,
\label{rie}
\end{equation}
is the spin-valued curvature tensor. Varying the action (\ref{action}) with respect to the auxiliary fields $ e $ and $ \chi $, we arrive at the following constraint equations
\begin{eqnarray}
H & = & -e^2 \frac{\delta S}{\delta e}=\frac{1}{2}\,
g_{\mu\nu}\dot{x}^\mu\dot{x}^\nu=0\,,
\label{h}
\end{eqnarray}
\begin{eqnarray}
Q =-i e \frac{\delta S}{\delta \chi}=e^a_\mu\dot{x}^\mu \psi_a =0\,.
\label{q}
\end{eqnarray}
From equation (\ref{h}) it follows that the particle moves along a null curve, while equation (\ref{q}) shows that its spin is spacelike.

\subsection{(Super)symmetries and Conserved Quantities}

The description of spacetime symmetries and the associated conserved charges in the motion of  spinning point particles in its most complete form  was given in \cite{gibbons,holten}. For our purpose in the following, we recall some ingredients of this description in the Hamiltonian  formalism. Introducing the basic phase-space variables $ (x^\mu\,, \Pi_\mu\,, \psi^a) $, where the covariant momentum
\begin{eqnarray}
\Pi_\mu &=& p_\mu + \frac{i}{2}\, \omega_{\mu a b } \psi^a \psi^b= g_{\mu\nu} \dot{x}^\nu\,,
\label{covmom}
\end{eqnarray}
we have the Hamiltonian in the form
\begin{equation}
 H=\frac{1}{2}\,g^{\mu\nu}\,\Pi_\mu \Pi_\nu\,.
\label{ham}
\end{equation}
The proper time evolution of any phase-space function $ J( x, \Pi, \psi) $  is determined  by the Poisson-Dirac bracket of this function with the Hamiltonian in (\ref{ham}). That is,
\begin{equation}
\frac{d J }{d\tau}= \{ J, H\}\,.
\label{evol1}
\end{equation}
The Poisson-Dirac bracket of two arbitrary phase-space  functions is  defined as follows
\begin{equation}
 \{ F, G\}={\cal D}_\mu\, F \frac{\partial G}{\partial \Pi_\mu}-\frac{\partial F}{\partial \Pi_\mu}\,{\cal D}_\mu G
 -R_{\mu \nu} \,\frac{\partial F}{\partial \Pi_\mu}\frac{\partial G}{\partial
 \Pi_\nu}+i(-1)^{\epsilon (F)}\,\frac{\partial F}{\partial \psi^a}\frac{\partial G}{\partial
\psi_a}\,\,,
 \label{pdb}
\end{equation}
where $ {\epsilon (F)} $ is either $\, 0 \,$ or $ \,1 \,$ depending on the Grassmann-even or odd parity of  $ F $. The phase-space covariant derivative is given by
\begin{equation}
{\cal D}_\mu F = \partial_\mu F +\Gamma_{\mu\nu}^\lambda \,\Pi_\lambda \,\frac{\partial F}{\partial \Pi_\nu}+\omega^{\ a}_{\mu\  b} \,\psi^b
\frac{\partial F}{\partial\psi^a}\,\,.
\label{pscovd}
\end{equation}
Clearly, for the vanishing Poisson-Dirac bracket in (\ref{evol1})
\begin{equation}
 \{J ,  H \}= 0\,,
\label{vanishpd}
\end{equation}
the set of the phase-space functions $ J( x, \Pi, \psi) $ can be thought of  as describing  all conserved quantities and  symmetries underlying the conservation. In this case,  substituting the expansion
\begin{equation}
J=\sum_{n=0}^\infty \frac{1}{n!} \,J^{(n)}_{ \mu_1 \dots  \mu_n}(x,\, \psi)\,  \Pi^{\mu_1} \dots \Pi^{\mu_n}\,
\label{consts}
\end{equation}
in equation (\ref{vanishpd}), one obtains the chain of equations
\begin{equation}
D_{(\mu_{n+1}} J^{(n)}_{\mu_1 \dots \mu_n)} + \frac{\partial J^{(n)}_{(\mu_1 \dots \mu_n}}{\partial\psi^a}\,\omega^{\ a}_{\mu_{n+1)}
   b }\,\psi^b=
    R_{\nu (\mu_{n+1}}
    J^{(n+1)\nu}_{\mu_1 \dots \mu_n)}\,,
    \label{chain}
\end{equation}
for different coefficients of the expansion $ J^{(n)}_{ \mu_1 \dots  \mu_n}(x,\, \psi) $. We use round brackets to denote symmetrization over the indices enclosed. It is easy to verify that these coefficients can be identified with Killing tensors of different ranks describing the symmetries of the action (\ref{action1}) (see for details \cite{holten}). The most simple solution of these equations for the second rank Killing tensor, $ K_{\mu\nu}= g_{\mu\nu} $, gives rise to the Hamiltonian (\ref{ham}), which generates the symmetries with respect to proper time translations. Meanwhile, the simple solution for the Grassmann-odd Killing vector, $ I^a=\psi^a $,  gives us  the supercharge  $ Q= \Pi_a \psi^a $  associated with the supersymmetry transformations of the form (\ref{supertrans}) with $ e=1 $ and $ \chi=0 $. Evaluating the Poisson-Dirac bracket of the supercharge  $ Q $ with the Hamiltonian (\ref{ham}) and with itself, we have the superalgebra
\begin{eqnarray}
\{ Q,  H\}& = & 0\,, \ \ \  \{Q, Q\}=-2 i H\,.
\label{superalg}
\end{eqnarray}
In obtaining the second  relation, we have used the cyclic identity for the curvature tensor in the expression
\begin{eqnarray}
 R_{\mu\nu} \frac{\partial Q}{\partial \Pi_\nu}& = &
 \frac{i}{2}\, R_{\mu[abc]}\psi^a\psi^b \psi^c=
 0\,,
 \label{cyclic}
\end{eqnarray}
where square brackets denote antisymmetrization over the indices enclosed. We note that the above symmetries and the corresponding dual symmetries (dual supercharge and chiral charge) along with the Hamiltonian itself are  of  generic symmetries in the sense that they are built in the  action (\ref{action1}) from the very beginning \cite{gibbons, holten}.  A natural question that arises  in this respect, is whether a classical system  described by the action (\ref{action1}) admits additional (nongeneric) supersymmetries. Clearly, the existence of these supersymmetries  will depend on the explicit form of the spacetime metric. In what follows, we discuss this question in the spacetime of general higher-dimensional rotating black holes.

\section{Nongeneric Supersymmetries}

The theory of spinning point particles in a curved spacetime possesses a new kind of supersymmetry if the spacetime metric admits a Killing-Yano tensor being a square root of the Killing tensor. In this respect, the new supersymmetry is an extension of the hidden symmetry generated by the Killing tensor.  The authors of work \cite{gibbons} were the first to construct explicitly a new supercharge, which depends on the second rank  Killing-Yano tensor and generates  this supersymmetry. In particular, they found that a spinning particle  in the Kerr-Newman  spacetime  acquires an additional worldline supersymmetry, corresponding to the Killing-Yano tensor discovered by Penrose and Floyd \cite{penrose}.

Recently, it was found that the most general rotating black hole spacetimes  described by the Kerr-NUT-(anti)de Sitter metrics \cite{clp} admit a closed conformal Killing-Yano (CKY) two-form that, in essence, is of a ``capsule"  for all hidden symmetries of these metrics \cite{fk1, fk2,fk3}. Building up the corresponding  exterior products of this CKY two-form and taking their Hodge duals one can generate a ``cascade" of Killing-Yano and Killing tensors, covering all spacetime dimensions. Here we use these results to construct new supercharges underlying the appearance of nongeneric supersymmetries for the spinning point particles in these spacetimes. In \cite{fk2} it was shown that with a suitable analytical continuation the Kerr-NUT-(anti)de Sitter metrics can be written in terms of the orthonormal basis one-forms  as follows
\begin{eqnarray}
ds^2 &= & \sum_{a=1}^D \sum_{b=1}^D \delta_{ab} e^a e^b= \sum_{\mu=1}^n \left(e^{\mu} e^{\mu} + E^{\mu} E^{\mu} \right) +\varepsilon  \omega \omega\,,
\label{orthoframe}
\end{eqnarray}
where $ n=[D/2]$ stands for the integer part of $ D /2 $ \,,  $ E^{\mu}= e^{n+\mu} \,\,,$  $ \omega= e^{2 n+1} $ and $ \varepsilon= D- 2n $, i.e. being $ 0 $ for even $ D $ and  $ 1 $ for odd $ D $.  For our purposes in the following,  we do not need the explicit expressions for these basis one-forms. We recall that  they can be found in \cite{fk2}.  In this orthonormal frame, the closed CKY two-form $ k $ existing in the Kerr-NUT-(anti)de Sitter metrics  acquires a skew-diagonal form which is given by
\begin{eqnarray}
k &= & \sum_{\mu=1}^n x_{\mu} e^{\mu}\wedge  E^{\mu}\,,
\label{cky2f}
\end{eqnarray}
where $ x_{\mu} $  are the eigenvalues corresponding to radial and latitude  directions in the canonical form of the spacetime metrics. The wedge products of the $j$-th power of this closed CKY two-form
\begin{equation}
k^{j}=k^{\wedge j}=k\wedge\dots \wedge k
\end{equation}
is again a  closed CKY form  and taking its Hodge dual one can  define the following nontrivial Killing-Yano $ (D-2j)$-form
\begin{equation}\label{towers}
f^{j}= * k^{j}, \ \ \  j=1,2,\dots \left[D/2-1\right]\,.
\end{equation}
The defining equations for the second rank closed CKY tensor have the form
\begin{eqnarray}
\label{ckyeqs1}
D_{c}k_{ab}&=& \eta_{ca}\xi_b-\eta_{cb}\xi_a \,, \\[2mm]
D_{[a}k_{bc]} & = & 0\,,
\label{ckyeqs2}
\end{eqnarray}
where
\begin{eqnarray}
\xi_a &=& \frac{1}{D-1} D_c k^c_{\,\,a} \,,
\label{trace}
\end{eqnarray}
while, the totally antisymmetric $ (D-2j) $ rank Killing-Yano tensor
\begin{equation}
f^{j}_{a_1 a_2  \dots a_{D-2j}}= f^{j}_{[a_1 a_2  \dots a_{D-2j}]}
\end{equation}
obeys the equation
\begin{equation}
D_{(a_1} f^{j}_{a_2) a_3  \dots a_{D-2j+1}}= 0\,.
\label{yano}
\end{equation}

Next, following the works of \cite{gibbons, cari},  we  consider the set of  supercharges
\begin{equation}
\Omega^j=  \Pi^{a_1}f^j_{a_1 a_2 \dots a_{D-2j}} \psi^{a_2}\cdots
    \psi^{a_{D-2j}} -
    \frac{i}{D-2j+1} D_{a_1} f^j_{a_2  \dots a_{D-2j+1}}
    \psi^{a_1} \psi^{a_2}\cdots
    \psi^{a_{D-2j+1}}\,\,,
\end{equation}
corresponding to the Killing-Yano tensors in various spacetime dimensions. It is straightforward to show that the  Poisson-Dirac bracket of these supercharges both with the Hamiltonian (\ref{ham}) and  the standard supercharge  $ Q $ vanishes. Evaluating first the  Poisson-Dirac bracket of $ \Omega^j $  with  $ Q $, we find that
\begin{eqnarray}
\{ Q, \Omega^j\}&=&-\left(\psi^a D_a \Omega^j + i \Pi^a \frac{\partial
\Omega^j}{\partial \psi^a}\right)
= -\left(D_a f^j_{a_1 a_2 \dots a_{D-2j}} \Pi^{a_1}
\psi^a \psi^{a_2}\cdots \psi^{a_{D-2j}}
\right. \nonumber \\[2mm]  & &  \left.
 -  \frac{i}{D-2j+1} D_a D_{a_1} f^j_{a_2 \dots a_{D-2j+1}}
    \psi^a\psi^{a_1}\psi^{a_2}\cdots
    \psi^{a_{D-2j+1}}
\right. \nonumber \\[3mm]  & &  \left.
+ D_{a_1}f^j_{a_2 \dots a_{D-2j+1}}
    \Pi^{a_1}\psi^{a_2}\cdots
    \psi^{a_{D-2j+1}}\right)\,.
    \label{qomega}
\end{eqnarray}
In obtaining this expression we have used the relation in (\ref{cyclic}). Substituting  in this expression the integrability condition
\begin{equation}
D_a D_{a_1} f^j_{a_2 \dots a_{D-2j+1}}= \frac{(-1)^{D-2j+1}}{2} \, (D-2j+1)\, R^b_{ \ a[a_1 a_2} f^j_{a_3\dots a_{D-2j+1}]b}
\label{integra}
\end{equation}
and taking into account equations (\ref{cyclic}) and (\ref{yano}), we see that it vanishes. Thus, we have
\begin{equation}
\{ Q, \Omega^j\} = 0\,.
\label{qomega1}
\end{equation}
The Jacobi identity  for two supercharges $ Q $, and $ \Omega^j $ implies that
\begin{equation}
\{ \Omega^j, H \} = 0\,.
\label{hamomega}
\end{equation}

Let us now consider a supercharge
\begin{equation}
\Upsilon = \Pi^a k_{ab} \psi^b\,,
\label{scharge}
\end{equation}
which corresponds to the second rank closed conformal Killing-Yano tensor. Evaluating the Poisson-Dirac bracket  of  $ Q $  with this supercharge,  we find  that
\begin{eqnarray}
\{Q,\Upsilon\} =
    - Q\, \psi_a \xi^a \,.
    \label{qups}
\end{eqnarray}
Clearly, this expression vanishes due to the constraint (\ref{q}). On the other hand, from the Jacobi identity, as in in the case of (\ref{hamomega}), it also follows that
\begin{equation}
\{\Upsilon, H\} = 0\,.
\label{hamups}
\end{equation}
Thus, both  $ \Upsilon  $ and  $ \Omega^j $ commute with the standard supercharge $ Q $ in the sense of   Poisson-Dirac brackets and
therefore, they are superinvariants.

Next, we are interested in vanishing  Poisson-Dirac bracket of $ \Upsilon  $ and  $ \Omega^j $ . It turns out that only for the supercharge  $ \Omega \equiv \Omega^{\left[D/2-1\right]} $  we have the vanishing Poisson-Dirac bracket. The novel feature of this supercharge is that  it corresponds to the second rank Killing-Yano tensor in all even spacetime dimensions, whereas  in all odd dimensions it depends on the third rank Killing-Yano tensor. We wish now to show that
\begin{equation}
\{ \Upsilon, \Omega \} = 0\,.
\label{upsom}
\end{equation}
For certainty, we  begin with  {\it even}  spacetime dimensions. In this case, we have the second rank Killing-Yano tensor
\begin{equation}
 f_{ab}\equiv f_{ab}^{(D-2)/2} = \frac{1}{(D-2)!}\,\varepsilon_{ab}^{\ \ \ \ d_1 d_2 \dots d_{D-3}\, d_{D-2}} k_{d_1 d_2}k_{d_3 d_4}\cdots k_{d_{D-3}\, d_{D-2}}\,,
 \label{2rkyt}
\end{equation}
where $ \varepsilon_{a\cdots}^{\ \ \ \ \ b\cdots} $ is the usual totally antisymmetric Levi-Civita  symbol, and the associated supercharge
\begin{equation}
\Omega= \Pi^a f_{ab}\psi^b
-\frac{i}{3}\,D_a f_{bc}\psi^a \psi^b\psi^c\,.
\label{prinscharge}
\end{equation}
Before calculating the Poisson-Dirac bracket (\ref{pdb}) for
$ \Upsilon  $ and  $ \Omega $, we first consider the last term in it, for which we have
\begin{equation}
\label{lt}
-i\frac{\partial \Upsilon}{\partial \psi^a}\frac{\partial \Omega}{\partial
\psi_a}=- \Pi_a k^{ab} \left(i \Pi^cf_{cb} +  D_b
f_{cd}\psi^c\psi^d\right)\,.
\end{equation}
Comparing equations (\ref{cky2f}) and  (\ref{2rkyt}) in even dimensions, we obtain the identity
\begin{equation}\label{id}
 k^{ab}f_{cb}= \frac{1}{D!!}\, \delta^{a}_{c}\,
\varepsilon_{d_1 d_2\dots d_{D-1} d_D}
k^{d_1 d_2}\cdots k^{d_{D-1} d_D}\,.
\end{equation}
Using this identity along with equations (\ref{h})\,, (\ref{q}) and (\ref{yano}) we transform equation (\ref{lt}) into the form
\begin{equation}
-i\frac{\partial \Upsilon}{\partial \psi^a}\frac{\partial \Omega}{\partial
    \psi_a}=  - \Pi^a D_c k_{ab}
    f^b_{\ d} \psi^c \psi^d\,.
\end{equation}
Performing now the similar calculations with the first two terms in (\ref{pdb}) for $ \Upsilon  $ and  $ \Omega $, and using the integrability condition (\ref{integra}), we find that
\begin{equation}\label{pdbf}
\{ \Upsilon, \Omega\}= - 3 \Pi^a D_{[c}k_{ab]}
f^b_{\ d} \psi^c \psi^d + i R_{h e b c}   f^h_{\ d} k^e_{\
a}\psi^{a} \psi^b \psi^c \psi^d\,.
\end{equation}
The first term in this expression vanishes due to equation (\ref{ckyeqs2}). On the other hand, using this equation one can construct the identity
\begin{equation}
D_a D_{[e} k_{bc]} f^e_{\ d} \psi^a\psi^b\psi^c\psi^d=0\,,
\end{equation}
which, in turn, results in
\begin{equation}
R_{a [e b}^{\ \ \ |h|}   k_{c]h} f^e_{\ d} \psi^a\psi^b\psi^c\psi^d= 0\,.
\end{equation}
This expression  along with (\ref{cyclic}) shows that the second term in (\ref{pdbf}) vanishes as well, thereby leaving us with (\ref{upsom}).

Turning now to  {\it odd} spacetime dimensions, we note that in this case the calculations for (\ref{upsom}) are entirely similar to those described above. We have the third rank Killing-Yano tensor
\begin{equation}
f_{abc}\equiv f_{abc}^{(D-3)/2} = \frac{1}{(D-3)!}\,\varepsilon_{abc}^{\ \ \  d_1 d_2 \dots  d_{D-4}\,d_{D-3}} k_{d_1 d_2}\cdots k_{d_{D-4} \, d_{D-3}}\,,
\label{3rkyt}
\end{equation}
and the supercharge
\begin{equation}
\Omega= \Pi^a f_{abc} \psi^b \psi^c
    -\frac{i}{4}\,D_a f_{bcd}\psi^a\psi^b\psi^c\psi^d\,.
\end{equation}
With equations (\ref{cky2f}) and  (\ref{3rkyt}) in odd dimensions, we find that
\begin{equation}
k^{ad}f_{cbd}= \frac{1}{(D-1)!!}
\,\,\delta^{a}_{[c} \varepsilon_{b]d_1 d_2\dots d_{D-2}\,d_{D-1}}
k^{d_1 d_2}\cdots k^{d_{D-2}\, d_{D-1}}\,.
\end{equation}
Using this identity and repeating all steps made above for the case of even dimensions, we again confirm  equation (\ref{upsom}). Thus, we have the set of three mutually commuting, in the sense of Poisson-Dirac brackets, supercharges  $ Q $,\,  $ \Omega $ and $ \Upsilon $.

\subsection{Extended superalgebra}

Though all the three supercharges commute  with each other however, unlike the case of  two standard  supercharges, as given in  equation (\ref{superalg}), the Poisson-Dirac bracket of two  supercharges $ \Omega $ as well as that of two supercharges  $ \Upsilon $, both
do not close on the Hamiltonian. Instead, they give rise to an extended, in some sense, unusual superalgebra. Following \cite{gibbons}, we find that in all even dimensions
\begin{equation}
\{ \Omega, \Omega \}=-2i \left(\frac{1}{2}\,K^{\mu\nu}_{(1)}
\Pi_\mu \Pi_\nu +I^\mu
    \Pi_\mu  +G \right)\,,~~~~~ even\,\, D\,,
\label{pdexo1}
\end{equation}
where  $ K^{\mu\nu}_{(1)} $ is a symmetric second rank Killing tensor given by
\begin{equation}
K^{\mu\nu}_{(1)}  = f^\mu_{ \ \lambda} f^{\nu\lambda}\,,
\label{killt1}
\end{equation}
$ I^\mu  $ is a spin-valued Killing vector,
\begin{equation}
    I^\mu =i \left(f^\nu_{ \ a} D_\nu f^\mu_{\ b} +
    f^{\mu\nu} D_\nu f_{ba}\right)\psi^a \psi^b\,,
\end{equation}
and the Killing scalar
\begin{equation}
G= -\frac{1}{4} \left(R_{\mu\nu ab} f^\mu_{\ c} f^\nu_{\ d} + 2 D_a f_b^{\ \mu} D_\mu f_{cd}\right) \psi^a \psi^b \psi^c \psi^d\,.
\end{equation}
We note that in odd spacetime dimensions  $ \Omega $ is Grassmann-even and therefore, its Poisson-Dirac bracket with itself identically vanishes
\begin{equation}
\{ \Omega, \Omega \}= 0 \,,~~~~~ odd \,\, D\,.
\label{pdzero}
\end{equation}
Similarly, for the Poisson-Dirac bracket of two  supercharges   $ \Upsilon $, we have
\begin{equation}
\{ \Upsilon, \Upsilon \}=-i K^{\mu\nu}_{(2)} \Pi_\mu \Pi_\nu\,,
\label{pdexo2}
\end{equation}
where the Killing tensor $ K^{\mu\nu}_{(2)} $ is given by
\begin{equation}
K^{\mu\nu}_{(2)} = k^\mu_{ \ \lambda} k^{\nu\lambda}\,.
\label{killt2}
\end{equation}
We recall that  in our constructions we use the constraint (\ref{h}), which corresponds to the null trajectories.

\section{The Myers-Perry metrics}

As we have emphasized above, our construction of the nongeneric supercharges refers to the most general rotating black hole spacetimes  described by the Kerr-NUT-(anti)de Sitter metrics \cite{clp}, though it does not require the explicit form  of these metrics. Below, for some illustrations of the results obtained in the previous section, we consider the general higher-dimensional Myers-Perry metrics given by
\begin{eqnarray}
ds^2 & = & -dt^2 +\frac{U dr^2}{V-2M} + \frac{2M}{U}\left(dt-\sum_{i=1}^n
 a_i\mu^2_i\, d\phi_i \right)^2 \nonumber
\\[3mm] &&
+ \sum_{i=1}^n \left(r^2+a^2_i\right)
 \left(\mu^2_i\,d\phi^2_i+  d\mu^2_i\right) + \varepsilon r^2 d\mu^2_{n+\varepsilon}\,\,,
\label{mpmetrics}
\end{eqnarray}
where the metric functions
\begin{eqnarray}
V=r^{\varepsilon-2}\prod_{i=1}^n \left(r^2+a^2_i\right)\,,~~~~~ \frac{U}{V} = 1-\sum_{i=1}^n
    \frac{a_i^2\mu_i^2}{r^2+a_i^2}\,\,,
\label{metfunctions}
\end{eqnarray}
and the latitude coordinates obey the relation
\begin{eqnarray}
\sum_{i=1}^n \mu^2_i + \varepsilon r^2 d\mu^2_{n+\varepsilon} &= & 1\,,~~~~~~~n= \left[(D-1)/2 \right]\,.
\label{dircos}
\end{eqnarray}
We note that here $ n= \left[(D-1)/2 \right] $, and
$ \varepsilon=1 $ for even and  $ \varepsilon=0 $ for odd spacetime dimensions. In \cite{fk1} it was shown that these metrics admit  the second rank closed conformal Killing-Yano tensor, which has the following form
\begin{equation}
k= \sum_{i=1}^n a_i \mu_i d\mu_i \wedge \left[a_i
dt - \left(r^2+ a_i^2\right) d\phi_i\right]+ r dr \wedge \left(dt -\sum_{i=1}^n a_i \mu_i^2 d\phi_i \right)\,,
\label{mppckyt}
\end{equation}
while, the associated Killing tensor is given by
\begin{eqnarray}
K^{\mu\nu} &=& \sum_{i=1}^n \left[a_i^2(\mu_i^2- 1)g^{\mu\nu}+ a_i^2
\mu_i^2 \delta^\mu_t \delta^\nu_t
+ \frac{1}{\mu_i^2}\,\delta^\mu_{\phi_i}\delta^\nu_{\phi_i}\right] \nonumber\\[2mm] &&
+ \sum_{i=1}^{n-1+\varepsilon} \delta^\mu_{\mu_i}\delta^\nu_{\mu_i} -2Z^{(\mu}Z^{\nu)} -2 \xi^{(\mu}\zeta^{\nu)}\,,
    \label{mpkillt}
\end{eqnarray}
where
\begin{equation}
    Z=\sum_{i=1}^{n-1+\varepsilon} \mu_i \partial_{\mu_i}, \ \ \  \ \xi=\partial_t, \ \ \
    \ \zeta= \sum_{i=1}^n a_i \partial_{\phi_i}\,.
\end{equation}
It is interesting to compare these results  with the Killing tensors in (\ref{killt1}) and (\ref{killt2}) obtained from the  Poisson-Dirac brackets of the corresponding  nongeneric supercharges. We begin with the $ D=4 $ case. Then,  the metrics (\ref{mpmetrics}) reduce to the usual Kerr metric and the second rank Killing-Yano tensor in (\ref{2rkyt})  takes the form
\begin{equation}
 f_{\mu\nu} = \frac{1}{2}\,\varepsilon_{\mu\nu}^{\ \ \alpha \beta } k_{\alpha\beta}\,.
 \label{2rmpkyt}
\end{equation}
Now, it is easy to see that
\begin{equation}
K^{\mu\nu}_{(1)}= \frac{1}{2} \,g^{\mu\nu}- K^{\mu\nu}_{(2)}\,.
\end{equation}
With the constraint (\ref{h}), these two Killing tensors are just proportional to each other, resulting in only one constant of motion. This is in agreement with (\ref{mpkillt}) in $ D=4 $. In the five-dimensional case $ D=5 $  we have the only  Killing tensor $ K^{\mu\nu}_{(2)} $, which  given by the  Poisson-Dirac brackets of two supercharges $ \Upsilon \,$.  This also agrees with  (\ref{mpkillt}).

However, the situation is completely different in dimensions $ D \geq 6 $, where the  supercharges $ \Upsilon $  and $ \Omega $
define two independent Killing tensors. To demonstrate  this explicitly, let us consider $ D =6 $. Using equation (\ref{mppckyt}), it is straightforward to show that the Killing tensor $ K^{\mu\nu}_{(2)} $ in (\ref{killt2}) agrees with the $ D=6 $ limit of the expression (\ref{mpkillt}), while the Killing tensor $ K^{\mu\nu}_{(1)} $ defined by the Killing-Yano tensor
\begin{equation}
f_{\mu\nu}=\frac{1}{4!}\,\varepsilon_{\mu\nu}^{\ \ \alpha \beta \lambda \tau} k_{\alpha \beta} k_{\lambda \tau}\,,
\end{equation}
has the following compact form
\begin{equation}
K^{\mu\nu}_{(1)} = \frac{1}{36}\left(3K^\mu_{(2)\lambda} K^{\nu\lambda}_{(2)} - 2 k^{\,2} K^{\mu \nu}_{(2)}
+ k^{\mu}_{\ \lambda}k^{\nu}_{\ \tau}K^{\lambda \tau}_{(2)} \right)\,,
\end{equation}
where $ k^2= k_{\mu \nu}k^{\mu \nu} $. Thus,  we have explicitly shown  that the extended algebra of  the nongeneric  supercharges $ \Omega $ and $ \Upsilon $, discussed in the previous section, results in  two independent  Killing tensors in the spacetime of Myers-Perry black holes with $ D=6 $. Clearly, this is also true in all $ D>6 $ dimensions.

\section{Conclusion}

The existence of hidden symmetries generated by the Killing-Yano tensors in rotating black holes spacetimes  has fundamental significance for the study of properties of the black holes in various dimensions. It is the Killing tensor that lies at the root of the appearance  of  a `` hidden supersymmetry" in  the motion of pseudo-classical spinning point particles in the Kerr-Newman spacetime \cite{gibbons}. The hidden supersymmetry appears as an extension of the usual worldline supersymmetry of the spinning particles and unlike, supersymmetry of black holes in  supergravity, does not require any special relation between the physical parameters of the black holes.

In this paper, we have studied the hidden supersymmetries  for the model of the spinning point particles in the spacetime of higher-dimensional rotating black holes. Using the fact that the Kerr-NUT-(anti)de Sitter metrics, describing the most general rotating black holes in all higher dimensions possess the hidden symmetries, we have presented two nontrivial supercharges based on the corresponding Killing-Yano and conformal Killing-Yano tensors of these metrics.
Evaluating the  associated Poisson-Dirac brackets, we
have shown that these supercharges along with the generic supercharge of the spinning particle model constitute a set of three mutually commuting supercharges. On the other hand, the hidden supersymmetries generated by the nongeneric supercharges do not obey the standard algebra of the generic supercharge, but form an unusual extended algebra. The latter, in turn,  results in two independent Killing tensors in spacetime dimensions $ D\geq 6 $.  These Killing tensors along with the three commuting Killing vectors  and the spacetime metric itself
guarantee  a complete separation of  variables for the Hamilton-Jacobi equation in six dimensions. For these dimensions, our construction may also shed some light on the physical reason of separation of variables in the Dirac equation \cite{oota}. On the other hand, it is known that in both cases the separability  occurs in all higher dimensions \cite{oota, fkk}. Therefore, our construction raises a natural question: What are the new extra supercharges underlying the  separability in  all $ D\geq 8 $ spacetime dimensions. This is a challenging  task for future work.

\section{Acknowledgments}

The authors thank  Nihat Berker and Teoman Turgut for their  stimulating  encouragements and support.

\end{document}